\documentclass[pdflatex,sn-mathphys-ay]{sn-jnl}

\usepackage{graphicx}%
\usepackage{multirow}%
\usepackage{amsmath,amssymb,amsfonts}%
\usepackage{amsthm}%
\usepackage{mathrsfs}%
\usepackage[title]{appendix}%
\usepackage{xcolor}%
\usepackage{textcomp}%
\usepackage{manyfoot}%
\usepackage{booktabs}%
\usepackage{algorithm}%
\usepackage{algorithmicx}%
\usepackage{algpseudocode}%
\usepackage{listings}%
\usepackage[symbol]{footmisc}%
\usepackage{endnotes}

\begin{document}

\title[Recommender systems, representativeness, and online music]{Recommender systems, representativeness, and online music: a psychosocial analysis of Italian listeners}



\author*[1]{\fnm{Lorenzo} \sur{Porcaro}}\email{lorenzo.porcaro@uniroma1.it}
\author[2]{\fnm{Chiara} \sur{Monaldi}}\email{chiara.monaldi@gmail.com}

\affil[1]{\orgdiv{Department of Computer, Control and Management Engineering}, \orgname{Sapienza University of Rome}, \orgaddress{\street{Via Ariosto 25}, \city{Rome}, \postcode{00185}, \country{Italy}}}
\affil[2]{\orgdiv{Studio Lago}, \orgaddress{\street{Via Suor Maria Mazzarello 43}, \city{Rome}, \postcode{00181}, \country{Italy}}}


\abstract{Recommender systems shape music listening worldwide due to their widespread adoption on online platforms. 
Growing concerns about representational harms that these systems may cause are increasingly part of the scientific and public debate, wherein music listener perspectives are oftentimes reported and discussed, but rarely contextualised through a psychosocial and cultural lens.
We address this gap by interviewing a group of Italian music listeners and analysing their narratives through Emotional Textual Analysis.
Our findings reveal that listeners often engage with platforms in routinized ways, yet lack a critical understanding of how recommender systems operate and experience a sense of detachment from algorithmic processes.
Moreover, while listeners perceive cultural and linguistic distinctions in music, their awareness of gender-related representational issues remains relatively limited.
Overall, results underscore the importance of integrating psychosocial insights with technical approaches in the design of trustworthy and culturally sensitive music recommender systems.
}




\maketitle
\newpage

\section{Introduction}\label{sec1}
Current music listening practices in digital environments are mediated by the inner mechanisms of online platforms, exposing listeners to an apparently seamless flow of music, whether through personalized playlists, next-song recommendations, infinite radios, or background music accompanying audio-visual content. 
The impact of these platforms has become a truly global phenomenon.
While services originating in Western markets were adopted early and spread rapidly, regional and local streaming and social media platforms are now emerging across different continents \citep{khalil2022digitality, ahmed2025african, qukaye2025shortvideo}.
Consequently, online platforms now constitute the central axis of everyday music listening worldwide, shaping both the global circulation of music and the ways listeners interact with it.

In this landscape, recommender systems, a particular class of information retrieval systems whose origins trace back to the end of the twentieth century \citep{Shoshana1992}, have assumed a central role as mediators between listeners and the vast music catalogues that are nowadays easily accessible. 
Their evolution, from early prototypes designed to replicate the mechanism of “word-of-mouth" \citep{Shardanand1995}, to the recent developments theorised as “companion" \citep{Karakayali2018}, highlights the sociotechnical nature of these systems \citep{Jannach2022}, which both shape and are shaped by people’s ways of enjoying music online.\endnote{Although recommender systems are often regarded as a subset of Artificial Intelligence (AI), and indeed represent some of its most successful real-world applications \citep{jannach_escaping_2020}, this work intentionally refers to them using the more precise term "Recommender Systems" to maintain clarity and avoid conflating technical specificity with broader contemporary AI discourse.}
Their importance has become even greater given the quasi-identical affordances that many online platforms, especially in Western markets, offer in terms of content availability and pricing. 
In this context, algorithmic and editorial curation, sometimes combined in what is described as \textit{algo-torial} curation \citep{Bonini2019}, have become key differentiating features for retaining listeners \citep{Hracs2021}.

Following the technological advancement and widespread adoption of recommender systems on online platforms, increased scrutiny has emerged regarding their influence on the music ecosystem \citep{Born2021, Hesmondhalgh2023}.
Despite a vast and ever-increasing body of literature on the impact of algorithmically driven recommendations, there is broad agreement that interdisciplinary enquiries in this area remain limited and are strongly needed \citep{Huang2023}. 
Such interdisciplinary enquiries are, we argue, especially needed to investigate the mechanisms through which recommender systems may contribute to representational harms, including processes of stereotyping, demeaning, erasure, alienation, and the over- or under-representation of socio-cultural groups \citep{Shelby2023}.
Under this lens, while the rise of online platforms has catapulted regional music artists onto the global stage, this increased visibility can expand opportunities while simultaneously posing systemic challenges, including the risk of cultural misrepresentation \citep{gomezgallegomunoz2025, ahmed2025african}.

In this article, we explore issues of representation in online music listening and the potential impact that recommender systems may have on them, focusing on the following research questions:

\begin{enumerate}
    \item[RQ1.] How do listeners perceive their relationship with online music platforms and music recommender systems?
    \item[RQ2.] How do listeners conceive of issues of representativeness when it comes to music platforms?
\end{enumerate}
Aiming to answer such questions, we interviewed a group of Italian listeners and analysed their narratives using a psychosocial interpretive model, the Emotional Textual Analysis (ETA) framework \citep{Carli2015}. 
We pay particular attention to the Italian music sector, motivated by its distinctive position.
Indeed, the growth of platform-based music consumption has coincided with a sustained predominance of local artists and repertoire \citep{damato2025italy}, making it a particularly valuable case for examining how recommender systems intersect with, adapt to, and potentially amplify existing cultural and market structures.

We argue that applying this psychosocial lens complements the state of the art in recommender systems technical research, which is predominantly centred around a cognitive-behaviorism standpoint \citep{Ekstrand2016, Said2025}. 
This predominance persists even in cases where approaches integrating psychological constructs and theories to model and predict user behavior, and to enhance recommendation processes, have been proposed \citep{Lex2021}.
This dominant paradigm for building and evaluating recommender systems limits the understanding of the overall user experience for several reasons.

First, although behavioral signals are easy to collect at scale, particularly from the perspective of platform owners, they are nonetheless sensitive to several factors such as user demographics, context, and the recency of interactions \citep{stray_building_2024}. 
As such, despite their apparent objectivity, these signals have well-documented limitations as proxies for representing underlying user preferences.
Second, there may be a mismatch between what users want and what they do \citep{Kleinberg2024}. 
Consequently, relying exclusively on behavioral data, such as user interactions with recommended music, may yield limited insights.
Third, research focused on engineering recommender system experiences rarely considers the role that \textit{algorithmic awareness}, i.e., the ability to recognize that an algorithmic system is in operation \citep{Gran2020}, and \textit{algorithmic knowledge}, i.e., the understanding of how such systems process information and how they may use data about the user when presenting content \citep{Hargittai2020}, can play in shaping user interaction.
Together, these concepts constitute the foundation of \textit{algorithmic literacy}, which also emphasizes critical understanding and informed application \citep{Laupichler2022}.

Through ETA’s psychosocial lens, we obtained a nuanced interpretation of Italian listener narratives that goes beyond traditional behavioral approaches, generating hypotheses about the relationship between listeners’ algorithmic literacy and their understanding of representational harms arising from human–recommender interactions.
Building on these insights, this study makes three key contributions.

First, by uncovering the emotional and symbolic dimensions of music listening through ETA, we offer a culturally situated interpretation of how listeners in the Italian context make sense of their interactions with algorithmic systems.
Second, the study advances the understanding of algorithmic literacy by mapping specific gaps between algorithmic awareness and critical knowledge among listeners.
In particular, we suggest that algorithmic literacy extends beyond technical competence to include the ability to critically engage with platform-mediated relations.

Third, by documenting the often surface-level reasoning surrounding representational harms, the findings reveal a disconnect between broader sociocultural debates and individual listener narratives.
We advance the hypothesis that, while listeners are attuned to linguistic and national cultural differences, often contrasting local identities with a globalized English-American standard, they tend to lack the interpretive frameworks needed to recognize more subtle algorithmic harms, such as gender stereotyping.
Collectively, these contributions bridge psychosocial approaches and technical research on music recommendation, underscoring the value of interdisciplinary perspectives for understanding algorithmic mediation in cultural consumption.

\section{ETA theoretical foundations}\label{sec2}
The Emotional Textual Analysis (ETA) framework is theoretically grounded in the conception of the unconscious on the basis of the first Freudian proposal:
the unconscious is understood as the emotional processing of contextual reality, and it has its own functioning different from the conscious. 
This proposal lays the foundation for the theories of the \textit{double reference principle} proposed by Fornari \citep{Fornari1979} and the \textit{bi-logic theory} introduced by Matte Blanco \citep{Blanco1975}, which together form the basis for understanding the dual nature of language and thought.

According to the double reference principle, language operates on two levels: 
the \textit{lexical-cognitive}, which involves the conscious, logical, and syntactic structure shared within a culture, and the \textit{symbolic-affective}, which encompasses the unconscious, emotional, and figurative meanings. 
The bi-logic theory complements this by describing two parallel modes of thought: 
the asymmetric, conscious logic that divides and rationalises, and the symmetric, unconscious logic that generalises and homogenises. 

Central to ETA is the concept of \textit{polysemy}, whereby words possess multiple potential emotional meanings.
\textit{Dense words}, rich in polysemy, carry significant emotional weight independently of context (e.g., “friend”, “explosion”), whereas non-dense words are more ambiguous and rely on their contextual relationships for meaning (e.g., “however”, “thing”). 
Emotionally dense words are not necessarily those that explicitly refer to recognizable emotions (e.g., “anger”, “fear”), rather, their emotional significance tends to be more implicit. 
Through etymological analysis, it becomes possible to uncover the emotional value embedded in words that are not directly linked to emotional vocabulary.
Verbs, in particular, play a crucial role in this analysis, as they help reveal underlying relational models.

Building on this, the ETA framework rests on three core concepts \citep{Carli2003}: emotional collusion, cultural repertoire, and local culture.
The notion of \textit{emotional collusion} captures the affective mechanism through which shared emotional understandings emerge within social interactions, progressively reducing polysemic meanings to shared symbolic processes.
This collective dynamic gives rise to the \textit{cultural repertoire}, a set of shared emotional–symbolic meanings that a social group employs to organize its collective relationship with a specific context.
The exchange and enactment of this repertoire constitute the \textit{local culture}, the broader socio-symbolic process through which members of a social group make sense of their experiences and regulate their behavior.

ETA has been applied in several contexts, including the experiences of caregivers providing in-home care to family members affected by dementia \citep{Caputo2021}, narratives of women undergoing assisted reproductive technology \citep{Langher2019}, family bonds and community distrust among younger generations \citep{Carbone2022}, representations of motherhood in detention \citep{Casellato2021}, and healthcare organizations providing services for adults with disabilities \citep{Bucci2021}.
In this work, this framework is applied, to our knowledge, for the first time, to analyze the relationship between people’s music listening habits and technology.
Having established the theoretical framework, we now turn to the methodology used to collect and analyze listener narratives.

\section{Methodology}\label{sec3}
The study followed a two-step procedure.
First, we interviewed a group of music listeners and transcribed the interviews to create a textual corpus for analysis.
Second, we applied ETA by identifying a set of dense words and analyzing their co-occurrence within textual units.
We then proceeded by interpreting dense words' relationships, finally synthesizing the emerging discourses in order to generate hypotheses and possible directions for future work.
Prior to the main study, three pilot interviews were conducted to refine the interview structure.
The pilot was conducted in May 2023, while the main study took place from September to November 2023.

\subsection{Participants}
Participants were recruited through the online crowdsourcing platform \textit{Prolific},\endnote{https://www.prolific.com} which allows prescreening to target specific populations.
Participants’ data were completely anonymized, and communication between participants and researchers was only possible within the platform. 

We did not aim to recruit participants with differentiated relationships to music, and the sole inclusion criterion was that participants be adult Italian citizens.
We focused on this group of listeners due to the researchers’ prior knowledge of and experience with the Italian cultural context.
It is important to note that individual characteristics are not central within ETA’s theoretical foundations.
Rather than describing how musical tastes or listening habits are distributed across a population, our goal is to identify underlying dimensions shaping how listeners interact with recommender systems and online platforms.
Accordingly, we did not collect personal data beyond age and gender, aiming to ensure a minimally balanced sample along these dimensions.

Twenty-one listeners were interviewed, all were born and resident in Italy and fluent in Italian.
Participants' ages ranged from 20 to 48 years ($M=30.8$, $SD=8.2$). 
For analysis purposes, age was grouped into three categories, 20–25 years, 26–35 years, and 36–48 years, reflecting early, mid, and later adulthood within our sample.
Gender was self-reported and treated as a binary variable (woman/man); in this sample, 46\% of participants identified as women.
We acknowledge that the sample is relatively small and culturally specific, and that the findings are exploratory in nature.
Indeed, they cannot be assumed to generalize to the broader population of Italian listeners or to other cultural contexts. 


\subsection{Interview settings and stimulus question}
We adopted an open interview format, presenting a stimulus question and allowing interviewees to freely associate their responses with it.
This format involves posing a single question that contextualizes both the interviewer–interviewee relationship and the research subject.
The choice of words used in the stimulus question is therefore pivotal and extensively discussed during the research design phase.
For this study, the designed stimulus question is:

\begin{quote}
    “We are interviewing a group of Italian citizens, and we are interested in the opinion of listeners, specifically with respect to the Italian music scene. In particular, we are interested in exploring the relationship between music listening and online platforms. Many studies are highlighting the influence of music recommendations on listening preferences. Through research, attention is being paid to representativeness, that is, how algorithms reflect musical realities. What are your thoughts on this?"\endnote{We report hereafter the original stimulus question in Italian, as it was asked to the participants: ``Stiamo intervistando un gruppo di cittadini italiani, ci interessa l’opinione degli ascoltatori, nello specifico rispetto alla realtà musicale italiana. In particolare, ci interessa esplorare il rapporto tra l’ascolto della musica online e le piattaforme di streaming. Molti studi stanno mettendo in luce l’influenza delle raccomandazioni musicali sulle preferenze di ascolto. Attraverso le ricerche si sta ponendo l’attenzione sulla rappresentatività, ovvero il modo in cui gli algoritmi rispecchiano le realtà musicali. Lei che ne pensa?"}
\end{quote}

During the interviews, everything said was considered relevant. 
The aim of this interview format is to explore the symbolizations that the interviewee offers through their response, similar to a psychological interview.
For this reason, the interviewer’s subsequent interventions did not steer participants’ responses and were limited to maintaining the conversational flow, for example by repeating the interviewee’s last words if silence suggested the interview might end prematurely.

This interviewing position requires careful training to appropriately support the interview process.
It is important to clearly understand the interviewer’s role and the research context that allows them to assume it.
This involves understanding the capacity in which interviewees are invited to participate, as well as why they may share the research objectives and be motivated to engage with the researcher.

\subsection{Corpus creation}
In this study, individual interviews lasted between 12 and 26 minutes ($M=17.0$, $SD=3.9$), for a total of 5 hours and 57 minutes. 
The interviews were automatically transcribed using \textit{Whisper} \citep{Radford2023},\endnote{https://github.com/openai/whisper} an open-source general-purpose speech recognition model, and then manually checked and cleaned.
The interviews were then compiled into a single corpus, reflecting the analytic approach in which individual interviews are not treated as relevant units, but rather the textual units derived from them.

To define textual units from the aural data, the organization of the transcriptions was primarily guided by prosodic features and discourse markers observed during the interviews.
Specifically, a new textual unit was identified when a significant pause, a shift in intonation, or a distinct turn-at-talk marked the end of a cohesive thematic segment.
These resulting textual units represent the final segments of discourse within which symbolizations related to the research object emerge. 
As such, textual units may be understood as a single, complete thought, argument, or thematic point articulated by the interviewee, composed of one or more syntactically linked clauses that are structurally and semantically self-contained.
Accordingly, the transcriptions were organized into final textual units and cleaned of the interviewer’s accompanying interventions.
An example of a textual unit from our corpus is reported hereafter:

\begin{quote}
    ``So, I think algorithms can help discover new... maybe new bands, new bands similar to the kinds I already listen to, as such I see it as something very useful, I mean, I might be used to listening to, say, three or four bands, and thanks a bit to real-time analysis of my listening habits, I get suggested things that are probably similar, or that people who listened to the same tracks I was listening to found enjoyable. Some of them can actually be really interesting, others not so much... but that’s a normal process."\endnote{We report hereafter the original quote in Italian: ``Allora penso che gli algoritmi possono aiutare a scoprire nuovi... magari nuovi gruppi, nuovi gruppi simili a tipologie che già vengono ascoltati, per cui lo vedo come una cosa molto utile, nel senso che io magari sono abituato ad ascoltare che ne so tre o quattro gruppi musicali e grazie un pochettino anche all'analisi in tempo reale dei miei ascolti, mi vengono proposte cose che probabilmente sono simili o che persone che ascoltano i brani che stavo ascoltando magari hanno trovati piacevoli, alcuni possono essere anche interessanti veramente, altri no... questo è un processo normale."} 
\end{quote}

Subsequently, we further preprocessed the transcriptions by removing punctuation and stop words, tokenizing the remaining text using \textit{NLTK} \citep{bird2009},\endnote{https://www.nltk.org} and lemmatizing every token using \textit{Simplemma} \citep{barbaresi2024}.\endnote{https://github.com/adbar/simplemma}
Some words considered significant with regard to the research context have been modified and made polyrhematic, that is, they were transformed into indivisible units whose overall meaning is autonomous from individual constituents, e.g., “music genre” has been replaced with “music-genre”.  
The processed corpus comprises 107 textual units totaling 13,827 words, with individual textual units ranging from 23 to 684 words ($M=129.2$, $SD=94.6$).

\subsection{Analysis procedure}
The analysis began with the identification of dense words across the entire interview corpus.
To this end, we constructed a vocabulary of all words used by the interviewees and selected dense words based on two criteria: (i) their frequency in the corpus, excluding words occurring fewer than ten times overall, and (ii) their relevance to the research questions, as determined through discussions within the research team.
The interpretation of dense words is guided by the hypothesis that their co-occurrence within textual units highlights the emotional-relational processes in the discourse.

We employed Multiple Correspondence Analysis (MCA) \citep{le2010} to uncover patterns of association between dense words and textual units. 
MCA is a dimensionality reduction technique whose output consists of low-dimensional embeddings of categorical variables, here, dense words and textual units.
These variables can be visually represented on a factorial plane using scatter plots, where the proximity between points reflects the strength of association among variables.

Additionally, we included two supplementary variables, self-declared gender and age group, to examine associations between interviewees’ demographic characteristics and the textual units they produced.
In practice, we created a contingency table with the dense words and the supplementary variables as columns, and textual units as rows.
This table served as input for the MCA, which was performed using the \textit{Prince} library \citep{Halford2024}.\endnote{https://maxhalford.github.io/prince}

Subsequently, we used the low-dimensional MCA representation as input for cluster analysis. 
Clustering was performed using the greedy K-means++ algorithm implemented in \textit{Scikit-learn} \citep{Pedregosa2011},\endnote{https://scikit-learn.org} with Silhouette analysis used to determine the optimal number of clusters.
Once the clusters were identified and positioned within the factorial plane, we proceeded with their interpretation.
Cultural repertoires are conceptualized under the hypothesis that the identified clusters organize cultural patterns that cut across individual interviewees’ speech.
As a result, each cluster constitutes a specific cultural repertoire in terms of ETA shared among several interviewees.
We analyzed the repertoires in order of prevalence, starting with those covering the largest proportion of textual units.

The first level of the repertoire analysis consists of interpreting the relationships between its dense words. 
This involved first identifying the most specific dense words within each repertoire.
To this end, we conducted a two-sided Pearson’s $\chi^2$ test with Yates’s continuity correction for each dense word, implemented in \textit{Scipy} \citep{virtanen2020},\endnote{https://scipy.org} where the null hypothesis assumed that the frequency of a dense word did not differ between textual units inside and outside a given repertoire.

Similarly, we conducted additional analyses to examine associations between demographic variables and cluster membership.
We conducted $\chi^2$ tests\endnote{We applied Fisher’s Exact test where necessary to ensure statistical validity across small samples.} to examine the relationship between gender and cluster membership, between age group and cluster membership, and additionally conducted post-hoc analyses to identify the specific categories driving the overall significance.
Through these analyses, we identified the dense words and demographic variables that most strongly characterize each cultural repertoire.
We then proceeded to the interpretive stage, considering only the five most characteristic dense words for each repertoire, ranked by their $\chi^2$ statistic, for which the null hypothesis was rejected ($\alpha = 0.05$).

The interpretation of each dense word began with an examination of its etymological or historical-cultural origins, as well as the lexical field to which it belongs (e.g., everyday register, codified language).
We interpreted the words in their original Italian form, however, we report their English translations, except in cases where the Italian wording is necessary to support the interpretation.
We then examined associative links by assessing whether a dense word appeared in the stimulus question and by connecting it to relevant literature on the topic discussed.
The sequential combination of dense words reduces their polysemy, thereby outlining the emotional meaning of the repertoire.

Finally, we examined the relationships among repertoires within the MCA factorial plane to interpret the overall dynamics characterizing the discourses present in the corpus.
The juxtaposition of repertoires was analyzed using psychological relational categories, interpreting cultures as polarized along specific dimensions (e.g., familiarity/strangeness; friend/foe).

It is important to note that our methodological choices were informed by a specific interpretive paradigm prioritizing conceptual clarity and theoretical grounding over purely statistical variance maximization.
Our primary objective was to uncover the fundamental antinomies and structural oppositions within the cultural repertoires. 
Drawing from established interpretive traditions in the psychological and social sciences, we argue that the most meaningful cultural differentiations often emerge along a small number of, typically bipolar, axes.

We also carefully considered the implications of applying additional non-linear dimensionality reduction techniques (e.g., t-SNE or UMAP) on top of the higher-order structure produced by MCA.
While such techniques are indeed valuable for uncovering complex manifold structures, they introduce an additional transformation layer, with the risk of compounding distortion. 
Thus, our decision to restrict the analysis and visualization to the first two MCA dimensions aims to preserve the closest possible alignment between the analytical dimensions and our theoretical interpretation of cultural antinomies. 
This approach helps minimize cumulative transformations that could obscure interpretability and ensures that our findings remain embedded in a transparent and theoretically coherent framework.

Having established our analytical framework, we now turn to the empirical findings. 
The following sections present the cultural repertoires that emerged from our analysis (Section \ref{sec4}), before proceeding to their qualitative interpretation (Section \ref{sec5}).

\section{Results}\label{sec4}
We identified 66 dense words, each occurring at least 10 times across the entire corpus.
After constructing a contingency table with dense words and supplementary variables as columns and textual units as rows, we fitted a two-dimensional MCA model to create the factorial plane. The two factors contributed 7.26\% and 5.84\% to the variance of the model, respectively.
The proportion of variance explained is low, indicating that the factors account for only a small fraction of the total variability in the data.
Low variance can make interpreting the MCA plot or understanding relationships among categorical variables more difficult. 
Patterns may be less pronounced, and distinctions between clusters may be less clear.
The results of the cluster analysis helped us better understand the relationships between dense words, supplementary variables, and textual units.
We selected $K=4$ clusters based on Silhouette analysis using candidate values $K = [2, 3, 4, 5]$. 
The algorithm was run ten times with different centroid initializations, and the run with the lowest inertia was retained as the final solution.

After assigning dense words and demographic variables to one of the four clusters, we performed a two-sided Pearson’s $\chi^2$ test with Yates’s continuity correction ($N = 13,827$, the total number of words in the corpus, $\alpha = 0.05$).
The $\chi^2$ statistics and p-values for all dense words are reported in the Supplementary Information, while Table \ref{tab1} lists only those selected for interpretation.

Additional analyses of demographic variables revealed significant associations between (i) gender and cluster membership ($\chi^2 = 18.67$, $p < 0.001$) and (ii) age group and cluster membership ($\chi^2 = 15.92$, $p = 0.014$).
Post-hoc analysis revealed a significant overrepresentation of male participants’ narratives in Cluster 2 ($p<0.001$), however, given the small sample size, this finding should be interpreted cautiously.
We discuss in Section \ref{sec6} our preliminary hypothesis regarding this overrepresentation, while noting that further research is needed to explore in greater depth the influence of specific demographic factors on these understandings, an important aspect, though not central to the present analysis.

Although the first two MCA dimensions account for a relatively small proportion of total variance, this is not a limitation within our interpretive framework.
ETA prioritizes uncovering symbolic-emotional structures and cultural antinomies embedded in discourse, which are often manifested across subtle patterns rather than dominating variance. 
Low variance in MCA does not imply trivial associations, rather, it reflects the diffuse and polysemic nature of natural language, in which multiple overlapping meanings coexist.
Cluster robustness was further ensured through repeated runs with varied initializations and by selecting the optimal number of clusters via Silhouette analysis.
Combined with ETA’s theoretical grounding, these procedures give confidence that the identified cultural repertoires meaningfully capture shared emotional–symbolic patterns in the corpus. 
They support interpretive validity despite the exploratory nature and low variance of the data.

\begin{table}[h!]
\caption{Summary of Cultural Repertoires. For each repertoire, the following information is reported: (i) coverage of textual units, and (ii) dense words with their corresponding $\chi^2$ statistic and p-value. Only dense words with p-value $<0.05$ were included in the analysis. For Cultural Repertoire 2, only four dense words meet this criterion.}\label{tab1}
\begin{tabular}{@{}llllll@{}}
\toprule
\multicolumn{3}{c}{\textbf{Cultural Repertoire 1 (Familiarity)}} & \multicolumn{3}{c}{\textbf{Cultural Repertoire 2 (Detachment)}} \\
\multicolumn{3}{c}{\textit{textual units: 49.53\%}} & \multicolumn{3}{c}{\textit{textual units: 18.69\%}} \\
\midrule
dense word&  $\chi^2$ & p-value & dense word&  $\chi^2$  & p-value \\
\midrule
Spotify & 27.27 & $<0.001$ & Algorithm & 35.28 & $<0.001$  \\
Influence & 20.54  & $<0.001$ & Ranking & 7.12 & 0.007  \\
TikTok & 19.13  & $<0.001$ & Piece & 5.54 & 0.018 \\
Platform  & 16.62  & $<0.001$ & To create & 4.26 & 0.038\\
Playlist & 16.45 & $<0.001$ & &   &   \\
\midrule
\multicolumn{3}{c}{\textbf{Cultural Repertoire 3 (Distinction)}} & \multicolumn{3}{c}{\textbf{Cultural Repertoire 4 (Representation)}} \\
\multicolumn{3}{c}{\textit{textual units: 20.56\%}} & \multicolumn{3}{c}{\textit{textual units: 11.21\%}} \\
\midrule
dense word&  $\chi^2$ & p-value & dense word&  $\chi^2$  & p-value \\
\midrule
English   & 35.94  & $<0.001$ & Man  & 121.82  & $<0.001$\\
Italian   & 22.58 & $<0.001$ & To represent & 94.97  & $<0.001$\\
American   & 19.32  & $<0.001$ & Woman  & 87.75 & $<0.001$ \\
Band   & 18.01  & $<0.001$ & Representativeness   & 58.17 & $<0.001$\\
Singer-songwriter & 17.02  & $<0.001$ & Difference  & 33.15 & $<0.001$\\
\botrule
\end{tabular}
\end{table}

\begin{figure}[h!]
    \centering
    \includegraphics[width=0.91\linewidth]{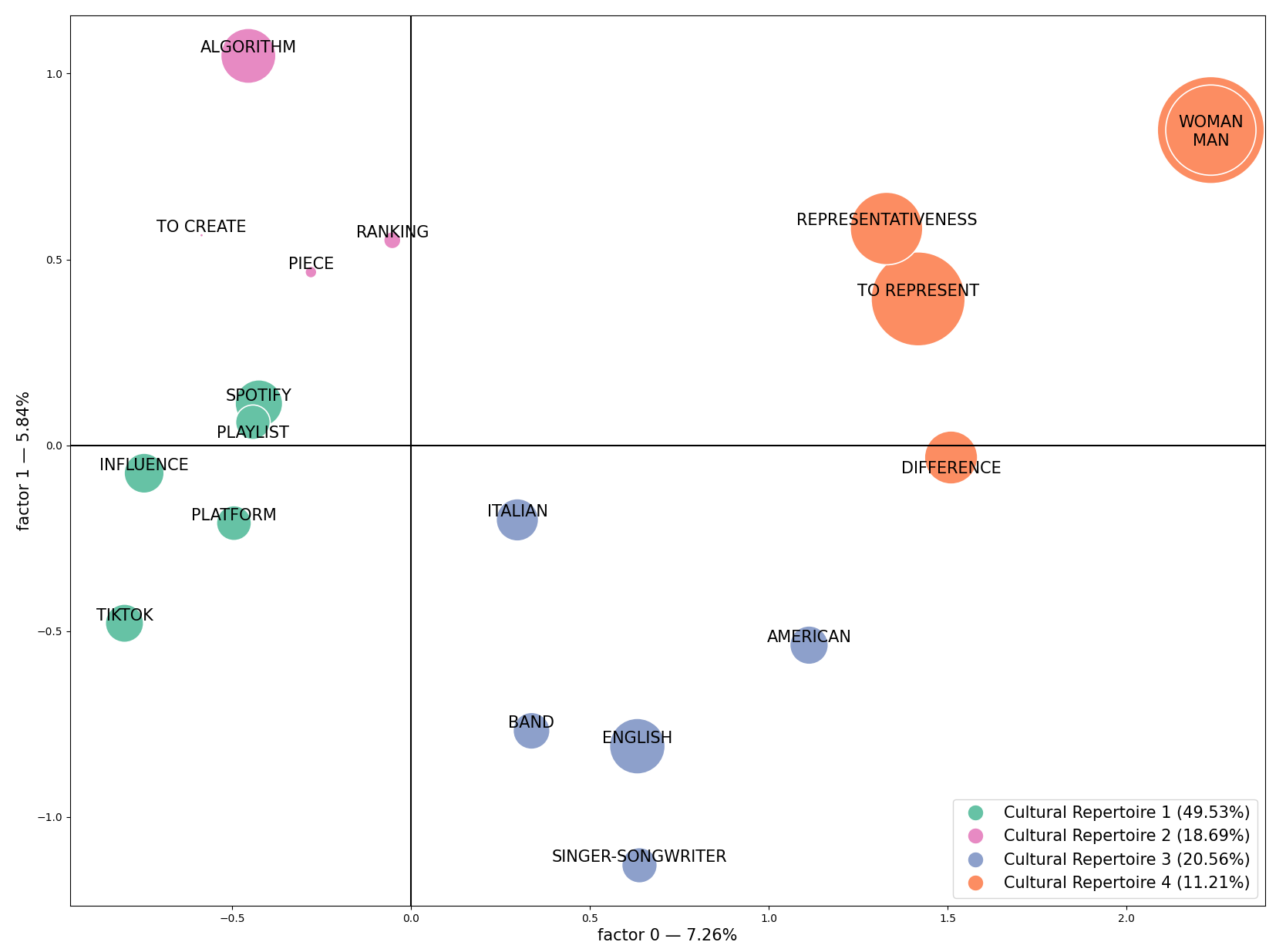}
    \caption{Visualization of cultural repertoires constructed using MCA and cluster analysis. Each point represents a dense word from Table \ref{tab1}, whose size reflects the $\chi^2$ statistic and whose color indicates the assigned cultural repertoire.}
    \label{fig:1}
\end{figure}

\section{Interpretative analysis}\label{sec5}
Figure \ref{fig:1} presents the four cultural repertoires as distributed across the factorial plane.
The x-axis separates technological discourses (Cultural Repertoires 1 and 2, left) from cultural discourses (Cultural Repertoires 3 and 4, right).
The y-axis separates terms used in familiar narratives (Cultural Repertoires 1 and 3, bottom) from more formal terms (Cultural Repertoires 2 and 4, top).

\subsection{Cultural Repertoire 1: familiarity between listeners and online platforms}

This repertoire is the most prominent, encompassing nearly half of the textual units from the interviews (47.66\%).
Among the dense words, two online platforms emerge, “Spotify” and “TikTok”, neither of which is mentioned in the stimulus question.

Spotify is the most popular music streaming service in Italy, and interviewees tend to equate online platforms with it.
TikTok, appearing lower than Spotify in the repertoire, highlights the role of social media in music discovery. 
On TikTok, music is associated with video content, sometimes for promotional purposes \citep{arantxa2023}, reshaping music circulation and supporting new forms of intercultural interaction \citep{li2025popular}.
The inclusion of specific platform names points to the emotional depth of these associations.
As observed by Siles and colleagues \citep{Siles2020}, when participants were asked to illustrate their connection to Spotify, it becomes clear that this relationship transcends mere utility. 
It embodies an interaction with a distinctly characterized technology.

The term “Influence,” second in the repertoire, carries substantial emotional weight, alluding to the intricate dynamics of the listening experience.
Present in the stimulus question, its etymology, from Medieval Latin “influentia,” derived from “influere,” meaning “to flow in”, hints at an action embedded within another.
Deconstructing this association reveals three entities: the platform, the music (or artist), and the listener, consistent with the multistakeholder recommender system literature \citep{abdollahpouri2021multistakeholder}.

We conceptualize “Influence” as a cyclic interaction: the platform shapes listeners’ habits \citep{ahmedelsheikh2025}, while listeners’ preferences simultaneously inform the platform’s algorithm \citep{avilatorresbeer2025}, a dynamic known as a \textit{feedback loop} in the recommender system literature \citep{Schmit2018}.
Emotionally, “Influence” embodies a nuanced duality, suggesting both guidance and subtle coercion and echoing the essence of recommendation in online platforms \citep{Seaver2019}.
This notion of influence implies a relationship with an entity loaded with emotional resonance. 
Moreover, it appears that interactions with platforms are characterised by a sense of customization, on the one hand, due to the personalization towards the listener, and on the other hand, due to the characterization of the platforms as personified \citep{Siles2020}. 

Other words in the repertoire further affirm its thematic focus on the symbiotic relationship between the listeners and the online platforms. 
“Platform", also part of the stimulus question, derives from the French “plate-forme", and contains the notion of a flat surface. 
Its presence in the stimulus question underscores its significance, particularly within the context of “online platforms", indicative of a sophisticated technological lexicon.  
However, its utility transcends mere technical jargon, mirroring a broader connotation of inclusivity and accessibility in the digital landscape. 
Compared to the term “service", which distinguishes between those who provide the service and those who use it, the term “platform" evokes a non-hierarchical dimension, something of which one may be a part \citep{Meier2019}.
Conjuring a spatial metaphor, this last dense word conveys a sense of egalitarianism, positioning everything and everyone on a level playing field.

The identification of “Playlist” as the last word in the repertoire underscores its pivotal role in both promoting and consuming music. 
Indeed, playlists serve a dual function: as a vehicle for artist promotion and dissemination, and as a mechanism for listener organisation, exploration, and personalization \citep{Aguiar2021}. 
Curiously, three of the first five words in this repertoire are of non-Italian origin, even though they are now part of everyday language in Italy.
This aspect prompts reflections on the emotional resonance inherent in technical terminology and its correlation to this particular repertoire. 

Having examined how listeners relate to platforms through familiar terms, we now turn to their conceptualizations of the algorithmic mechanisms underlying these platforms. 
This perspective reveals a markedly different emotional register.

\subsection{Cultural Repertoire 2: algorithms and the listeners detachment}\label{sec:CR2}
This cultural repertoire is characterised by its sparse lexicon, constituting a small fraction (18.69\%) of the textual units. 
Its close kinship with the previous cultural repertoire is reflected in the nature of its dense words.

The repertoire opens with the word “Algorithm”, included in the stimulus question and serving as the core subject of this repertoire.
In the context of our interviews, we hypothesize that this term conveys a sense of detachment among the participants.
Indeed, the algorithmic components of listener-platform interactions remain largely unexplored. They represent a process that, while not fully understood by participants \citep{Alvarado2018}, is nevertheless experienced daily and, in turn, imagined \citep{Bucher2017}.
It is also important to note that in participants' narratives, the term “Algorithm” substitutes the more technical and less commonly used term “Recommender System”.
This corroborates the idea that the listeners may not be fully aware of the technicalities of the algorithms that govern the exposure to music on online platforms.

Similarly, the other words in this repertoire appear to convey a sense of formality and detachment.
The following word, “Ranking" (in Italian “Classifica”), derives from the verb “classify” which originates from the concept of “class". 
The etymology of “Ranking" is rooted in the notion of an organised and distinct group. 
Within our research, “Ranking" also denotes the hierarchical ranking of songs – often referred to as music charts. 
Notably, a nuanced contrast emerges between “Playlist”, part of the first repertoire and often crafted by listeners themselves, and “Ranking”, which is governed by predetermined quantitative criteria \citep{hakanen1998}.
Under this lens, a subtle aura of detachment surrounds the concept of “Ranking", in which listeners, relegated to passive consumption, stand apart from the creative process.

The next term, “Piece”, similarly carries a sense of formality.
“Piece" (in Italian “Brano”) is a noun derived from the Old French “braon", meaning “piece of meat", originally connotating a fragment of something. 
It is important to consider the emotional distinction expressed in Italian between “Piece" and “Song" (“Brano” versus “Canzone”). 
In the music field, “Piece" carries a formal connotation, often referring to a segment of a larger work, whilst the word “Song" maintains a familiar and popular dimension \citep{fabbri1982}. 
In this repertoire, “Piece” appears to reflect the language of the platforms rather than that of the participants.
Tellingly, in translations from English to Italian used by online platforms the word “Song” is often replaced with “Brano" (e.g., translating “Liked song” as “Brani che ti piacciono”, in Italian literally “Pieces that you like”). 
Consistent with the other words in the repertoire, this lexicon appears to express a sense of detachment experienced in the relationship between the listeners and the algorithm.

The sole verb in this repertoire is “To create”, whose primary meaning is to bring something into existence from nothing, often in reference to a deity.
Another broader meaning indicates the act of producing, constructing, founding, inventing, or giving life to something that did not previously exist. 
Within this repertoire, the listening practices described do not involve a personalized experience but instead reflect a relationship with an algorithm symbolized as a distant and superior entity \citep{singler2020}.
The act of creating does not seem to occur within a reciprocal relationship between the listener and the algorithm, as found in the first repertoire while considering online platforms.

This contrast between familiar platform engagement and algorithmic detachment highlights a fundamental tension in participants' technological experiences. 
We now turn from this technological discourse to examine how participants articulate cultural differences in music.

\subsection{Cultural Repertoire 3: distinction in music culture and language}
Ranking second in terms of textual unit coverage (22.43\%), this cultural repertoire is highly polarized, addressing differences along linguistic and cultural dimensions, even if not directly related to representativeness or explicitly connected to online listening dynamics.
We interpret this repertoire as one of cultural opposition, highlighting the dichotomies “Italian” versus “English” (“American”) and “Singer-songwriter” versus “Band”.
These terms refer not only to languages but also to distinct musical genres, reflecting the broader contrast between local music culture and Western musical influences \citep{Born2000}.

The repertoire begins with the word “English" evoking notions of language and musical heritage. 
We associate this term with the dominant language of the Western music market from the post-war period onward \citep{Achterberg2011}.
The prominence of this word within the repertoire suggests the centrality of English language offerings within online platforms \citep{Page2023}. 
This repertoire appears to intersect with our stimulus question, particularly concerning the segment related to the listening of Italian music. 
Notably, the most prominent word in this repertoire is “English”, suggesting that defining Italian music, or music in Italian, implicitly involves comparison with English-language music.
Similarly, the term “American” introduces the theme of U.S. cultural influence in music production and promotion, highlighting the impact of American music on global audiences \citep{Colista1998}.

In contrast, the word “Italian” appeared in our stimulus question.
Given the interview context and the sample of native Italian speakers, we interpret this word as reflecting the participants’ cultural identity.
In our interviewing experience, respondents often began by stating that they “...did not listen to Italian music", associating it not simply with music originating from Italy or in the Italian language, but oftentimes with the Italian pop music genre \citep{Varriale2016}.
Thus, this dense word signifies linguistic identity while also warranting consideration in relation to Italian music.

Another term in this repertoire is “Band”, an English noun derived from the French “bande”, historically denoting membership in an organized group.
In the music field, it indicates a group of musicians, e.g., jazz band, a small-sized orchestral ensemble performing jazz music. 
It is worth mentioning that during the interviews, the participants mentioned the term “band” and not its Italian translation “banda (musicale)”. 
In this repertoire, which opens with “English”, the choice of such terms appears to evoke a mythical dimension of American music \citep{church2019}.
In fact, to describe a group of people making music together an English term is borrowed, even if the band does not necessarily originate from the UK or USA.

The final word of this repertoire, “Singer-songwriter” (Italian: “Cantautore”), is a term introduced in Italy in the 1960s to denote popular music performers who also wrote the lyrics for their songs \citep{Santoro2002}.
This designation emerged to categorise a distinct local musical genre. 
In the context of our inquiry into Italian music, it pertains specifically to Italian singer-songwriters, thus bearing significance in relation to the native language of our interviewees and connoting a sense of cultural affiliation. 

While cultural and linguistic differences readily emerge in participants’ narratives, questions remain regarding how they engage with other dimensions of representation. 
We now turn to the fourth repertoire, in which the theme of representativeness, explicitly raised in our stimulus question, emerges in relation to gender, though in terms markedly different from the cultural distinctions just discussed.

\subsection{Cultural Repertoire 4: exploring differences in gender representativeness}
The final cultural repertoire engages with the theme of representativeness, explicitly raised in the stimulus question.
The words in this repertoire are broad, referring to general categories, and notably, the theme of representativeness appears somewhat detached from the broader discourse on online music listening.
This is further indicated by the limited coverage of textual units (11.21\%), suggesting that participants devoted comparatively little attention to these topics during the interviews.
We note our deliberate choice to keep “To represent” and “Representativeness” distinct, preserving the distinction between the action and the conceptual noun. 
The verb tends to be more polysemous, while the noun is primarily used within a sociopolitical semantic context.

This repertoire opens with the word “Man", symbolising representativeness primarily in terms of gender roles. 
“Man” is perceived as a standard reference, connecting social issues to the overarching prominence of the male gender in the Italian language and culture.
However, the theme of gender seems somewhat lacking in contextualization. 

The first verb in this repertoire, “To represent” (from the Latin “repraesentare”, composed of “re-” and “praesentare”, meaning “to present”), carries a dual emotional resonance: on one hand, the act of making something present or visible, on the other, acting on behalf of or representing others.
We think about the verb in relation to the word "Man". 
Here, “Man” is extended to a generic class, and the notion of representation emerges as a reduction of this generalization.

The term "Woman" appears following the verb “To represent". 
“Woman” does not appear to function emotionally as the counterpart to “Man”, but rather as a subcategory arising from the reduction inherent in representation.
In the context of our research and online music listening, this evokes the issue of women’s under-representation in the music field \citep{Epps2020,Crider2022}.
We believe that such an issue arises concerning a subset of the general category. 
For example, playlists are often organized by gender rather than musical style, such that collections of female artists are grouped primarily based on gender membership rather than genre \citep{Werner2020}.

Following, we have “Representativeness”, a term present in the stimulus question. 
As argued by Chasalow and Levy \citep{Kyla2021}, ``Representativeness is a foundational yet slippery concept which presents multiple meanings [...] from typical or characteristic, to a proportionate
match between sample and population, to a more general sense of accuracy, generalizability, coverage, or inclusiveness".
In the interviews, it is linked to gender -- not unsurprisingly considering that gender representativeness is a quite common topic of interest also outside the music sector \citep{mazieres2021,Ulloa2024} -- however without broader cultural considerations.
%

Finally, the word “Difference” evokes two potential relational modes, one indicating diversity between entities at the same level, and another marking a gap or inequality.
Based on the analysis of the previous dense words in this cultural repertoire, gender differences are presented as a gap in the context of our research.

In sum, the limited engagement with gender representativeness, reflected in both sparse textual coverage and broad, generic vocabulary, contrasts with the more pronounced cultural distinctions observed in the previous repertoire.
Having interpreted all four cultural repertoires, we now turn to consider the broader implications of these findings.

\section{Discussion}\label{sec6}
The four cultural repertoires provide complementary insights into our research questions, revealing two fundamental tensions in listeners’ narratives.
First, concerning the relationship with platforms and recommender systems (RQ1), listeners engage with them in routinized, everyday ways, yet experience a sense of detachment from the underlying algorithmic processes, highlighting a dual dynamic between familiarity and limited understanding.
Second, regarding conceptions of representativeness (RQ2), listeners recognize cultural and linguistic distinctions in music, yet their awareness of gender-based disparities is superficial and fragmented, indicating limited engagement with broader representational concerns.
We now examine these tensions and their implications for algorithmic literacy and representational awareness.

\subsection{Bridging the literacy gap: from familiarity to critical knowledge}
What emerges from the interpretation of the overall dynamics of the Cultural Repertoire 1 and 2 is the co-existence of two modes of relation between listeners, online platforms, and recommender systems. 
On one hand, listeners perceive interactions with platforms in familiar terms, integrated seamlessly into their daily listening routines.
The relationship between listeners and platforms is mediated by reciprocal influence, with both actors shaping and being shaped by each other’s actions.
In this regard, influence is not emotionally characterised as positive or negative, but assumed as a core element, needed in order to maintain the integrity of this relation.

On the other hand, still the algorithmic components constituting online platforms' recommender systems seem to pertain to a mythological and opaque sphere in listeners’ thoughts, unknown entities which may define how people experience music online. 
Viewed through this lens, the narratives reveal a unidirectional, asymmetrical relationship, in which the algorithm creates while the listener primarily consumes.
It is important to remark on this difference between the role of online platforms versus algorithmic-driven recommender systems.
The former is a place wherein listeners have the power to contribute, the latter is an entity that, in the listeners’ imagination, is beyond the range of influence.

We also note that the Cultural Repertoire 2, where we hypothesize a sense of detachment between listeners and algorithms, is the only one in which we observed a significant overrepresentation of male participants’ narratives.
We interpret this difference as potentially reflecting persistently low algorithmic literacy in Italian education and professional contexts, a phenomenon disproportionately affecting women and girls.\endnote{Among Italian graduates in 2022, the share of ICT graduates was particularly low, at only 1.5\% compared to the EU average of 4.5\%, and of these, just 0.3\% were women \citep{EC2024Italy}.
Similarly, in 2024, Italy recorded one of the lowest proportions of ICT specialists in total employment (4.0\%), below the EU average of 5.0\%, with women accounting for only 17.1\% of ICT specialists, compared to 19.5\% in the EU \citep{Eurostat2025ICT}.}

Consequently, we hypothesize that the detachment observed predominantly among male participants, potentially due to low literacy levels, may also highlight a subtle but important aspect: 
among female participants, technical vocabulary and understanding of the systems mediating online music listening appear even more limited.

These insights carry practical implications for the design of recommender systems in online platforms. 
First, developing a shared technical vocabulary emerges as a priority, as its absence likely contributes to the sense of detachment.
Second, interfaces that more clearly explain the ``how'' behind recommendations, and not just the ``why'', could help listeners move from passive consumption to a more informed and active engagement. 
Similarly, fostering listener agency is crucial, as emphasizing how users’ actions shape recommendations could transform routine listening into literate engagement, bridging the gap between everyday use and algorithmic understanding.

\subsection{Addressing representational harms: beyond cultural dichotomies}

In parallel, the Cultural Repertoires 3 and 4 show how the concept of representativeness may be detached from the general discourse of online music listening. 
When solicited, representational issues are mostly connected by listeners with gender differences. 
Undoubtedly, the disparate impact of algorithmic curation on different gender identities continues to be a sensitive theme, as gender inequality is also a well-known issue in the music sector. 
Nonetheless, we observe superficial reasoning around this theme, reflected in the use of a broad and general vocabulary.
This leads us to think of an absence of complex and in-depth thought on potential representation harms caused by online platforms and their recommender systems.

On the contrary, cultural differences tend to emerge quite naturally when connected with linguistic diversity and nationality, in particular by contrasting English-American culture, the “global” from the perspective of an Italian citizen, with their own culture, the Italian scene. 
These two cultural discourses, gender and national-linguistic, are clustered in two different quadrants of the factorial plane, therefore somehow not directly connected in listeners' narratives. 
However, both occupy the right region of the plane, suggesting that these topics share common ground when considered under the broader umbrella of cultural differences in listening practices.

These observations suggest that, because listeners may not fully perceive gender disparities or other subtle representational harms, platforms cannot rely solely on user feedback to identify these issues. 
Music recommender systems therefore may require proactive auditing to ensure they do not reinforce alienating or narrow representations of socio-cultural groups. 
At the same time, involving listeners in these assessments can enhance both transparency and literacy, allowing them to better understand how their actions influence recommendations and to engage more actively with the system. 
Integrating psychosocial insights with traditional behavioral metrics can help reveal when algorithms inadvertently constrain listeners’ cultural horizons or perpetuate societal biases, enabling platforms to respond more effectively to users’ actual experiences and preferences.

\section{Limitations}
What we present hereafter has neither the ambition nor the aim of generalising our findings to a larger or more varied group of listeners.
Unlike quantitative approaches that aim to measure user experience objectively, ETA generates interpretative hypotheses grounded in symbolic patterns.
The outcomes are not statistically generalizable, rather, they provide insights into the emotional and relational structures embedded in discourse.
Moreover, focusing on Italian listeners may limit the generalizability of our findings.
Nevertheless, the methodology could be applied to other national or regional cultures, provided that local experts and scholars in music and psychosociology are involved.

Another important consideration for the scope and transferability of our findings concerns the socio-technical characteristics of our sample.
Indeed, we acknowledge that we did not systematically collect data on participants’ educational background, socio-economic status, or their experience and familiarity with music recommender systems and streaming platforms.. 
Given that digital competence and access to technology are stratified by class and education, these factors likely shape participants’ interactions with online platforms and their subsequent reflections.

Therefore, while our study identified several underlying structural dimensions of listener interaction among the participants, the prioritization and expression of these dimensions may differ in populations with lower or higher educational attainment, or with varying levels of exposure to recommender systems.
Future research is needed to test the transferability of these dimensions across different socio-economic strata and levels of user experience.

Lastly, although the open-ended stimulus question aligns with the interpretive aims of ETA, it may have constrained the emergence of specific reflections on recommender system mechanisms.
As an exploratory study, our goal was to identify potential directions for future research and to highlight possible interventions grounded in the socio-cultural interactions people engage in daily within online environments.

\section{Concluding remarks}\label{sec7}
Synthesizing the findings, we identify two main directions that warrant further investigation through more focused explanatory and empirical approaches.
The first direction concerns the role of algorithmic awareness, knowledge, and literacy in shaping the user experience.
Although recommender systems are nowadays part of the daily listening experience, and contribute to the development of a familiar relationship between listeners and platforms, the opaqueness of their functioning appears to limit the emergence of a critical meta-reflection on the terms of this relationship. 
This highlights a connection with the principles of transparency and explainability, widely regarded as pillars for constructing trustworthy systems \citep{Wang2023}.

Notably, familiarity with a technological artifact does not necessarily translate into an understanding of its logic, nor does it eliminate the sense of estrangement associated with algorithmic decision-making.
We question whether specific interventions, such as explaining why a recommendation is provided, can move beyond informing users and foster a more reflective understanding of the reciprocal influence between listeners and platforms.
In brief, to what extent is there an interest in making music listeners more aware not only of the functioning of recommender systems but also of the reciprocal relationship established through their interaction with online platforms?

The second, more concerning direction relates to representational harms and their connection with gender inequalities in the music sector.
We observe that gender, along with differences in language and nationality, two of the primary identity dimensions considered when studying the impact of music recommendations \citep{Hesmondhalgh2023}, emerge in listeners’ narratives, yet do not appear to be embedded within a broader reflection on diversity, non-discrimination, and fairness.
While under-representation of women and gender minorities in algorithmic recommendations has been widely explored in the literature, our findings suggest that awareness of such disparities, and of representational harms more broadly, remains limited among music listeners.

In conclusion, these findings indicate that challenges of transparency, explainability, and representation in music recommender systems are not solely technical, but equally cultural.
Addressing these challenges therefore requires not only system-level interventions but also a deeper understanding of how listeners conceptualize, interpret, and negotiate their relationship with algorithmic mediation.

\backmatter
\newpage
\theendnotes








\subsection*{Acknowledgements}
This work was done when Lorenzo Porcaro was at the European Commission's Joint Research Centre (JRC). The opinions expressed are those of the author(s) only and should not be considered as representative of the European Commission’s official position. This work is partially supported by the HUMAINT programme, Joint Research Centre, European Commission.

\subsection*{List of Figures}
\begin{itemize}
    \item[Figure 1.] Visualization of cultural repertoires constructed using MCA and cluster analysis. Each point represents a dense word from Table \ref{tab1}, whose size reflects the $\chi^2$ statistic and whose color indicates the assigned cultural repertoire.
\end{itemize}

\section*{Declarations}


\subsection*{Ethical approval}
The methodology followed in the study was subject to ethics review procedures defined by the European Commission's Joint Research Centre.

\subsection*{Informed consent}
Participants were fully informed about the voluntary nature of their participation, and they had the freedom to withdraw from the study at any point. 
They were also informed about their rights, including the right to access, correct, and delete their personal information. 
Furthermore, the information sheet outlined the research objectives, methodology, potential risks, and benefits. 
Informed consent was obtained from all participants, ensuring they agreed to participate with full awareness of the study's details and their rights.

\subsection*{Competing interests}
Chiara Monaldi's work has been funded by the European Commission's Joint Research Centre. She has received compensation as an external expert. Lorenzo Porcaro declares no potential conflict of interest.

\subsection*{Author contributions}
The authors contributed equally to this work. 

\subsection*{Data and code availability}
All data and code generated or analysed during this study are included in this published article and its supplementary information files.

\end{document}